\documentclass[apj]{emulateapj}
\usepackage{soul}
\usepackage{color}
\usepackage{amssymb}
\usepackage{graphicx}
\usepackage[normalem]{ulem}  % \sout{old text} for strikeout
\usepackage{natbib}

\newcommand{\midtilde}{\raisebox{-0.25\baselineskip}{\textasciitilde}}

\begin{document}

\title{Core-collapse supernova equations of state based on
neutron star observations}

\author{A. W. Steiner}
\affil{Institute for Nuclear Theory, University of Washington,
  Seattle, WA 98195}

\author{M. Hempel}
\affil{Department of Physics, University of Basel, 
Klingelbergstrasse 82, 4056 Basel, Switzerland}

\author{T. Fischer}
\affil{GSI, Helmholtzzentrum f\"ur Schwerionenforschung GmbH, 
Planckstra{\ss}e 1, 64291 Darmstadt, Germany \\
Technische Universit{\"at} Darmstadt, 
Schlossgartenstrasse 2, 64289 Darmstadt, Germany}

%\author{I. Sagert}
%\affil{Michigan State University, 1 East Lansing, Michigan 
%48824-1321, US}

\begin{abstract}
Many of the currently available equations of state for core-collapse
supernova simulations give large neutron star radii and do not provide
large enough neutron star masses, both of which are inconsistent with
some recent neutron star observations. In addition, one of the
critical uncertainties in the nucleon-nucleon interaction, the nuclear
symmetry energy, is not fully explored by the currently available
equations of state. In this article, we construct two new equations of
state which match recent neutron star observations and provide more
flexibility in studying the dependence on nuclear matter properties.
The equations of state are also provided in tabular form, covering a
wide range in density, temperature and asymmetry, suitable for
astrophysical simulations. These new equations of state are
implemented into our spherically symmetric core-collapse supernova
model, which is based on general relativistic radiation hydrodynamics
with three-flavor Boltzmann neutrino transport. The results are
compared with commonly used equations of state in supernova
simulations of 15 and 40~M$_\odot$ progenitors. We do not find any
simple correlations between individual nuclear matter properties at
saturation and the outcome of these simulations. However, the new
equations of state lead to the most compact neutron stars among the
relativistic mean-field models which we considered. The new models
also obey the previously observed correlation between the time to
black hole formation and the maximum mass of an $s=4$ neutron star.
\end{abstract}

\preprint{INT-PUB-12-033}

\maketitle

\section{Introduction}

% general intro to SNe
Core-collapse supernovae (SNe) are some of the most energetic events
in the Universe. They release several times $10^{53}$~erg in
neutrinos, the gravitational binding energy difference between iron
core and neutron star, and in case of SN explosions several
$10^{51}$~erg kinetic energy of the ejected material. The latter are
related to the standing accretion shock revival, which forms when the
super-sonically collapsing stellar core reaches nuclear matter
densities and bounces back. The shock wave initially propagates out of
the stellar core and thereby loses energy due to the dissociation of
infalling heavy nuclei and electron neutrino escapes when it crosses
the neutrinospheres. Several explosion mechanisms have been discussed,
i.e magneto-rotational by \citet{LeBlanc:1970kg}, dissipation of sound
waves by \citet{Burrows:2005dv}, and the standard scenario driven by
neutrino heating by \citet{Bethe:1985ux}.

% eos in SNe, general
The equation of state (EOS) is one of the critical and highly
uncertain microphysical inputs for modeling core-collapse supernovae.
The EOS in SN simulations has to handle several intrinsically
different regimes. For temperatures below about 0.5~MeV,
time-dependent nuclear reactions are important in determining the
nuclear composition. Above 0.5~MeV, nuclear statistical equilibrium
can be applied, where the dependence of the EOS can be reduced to the
three independent variables: temperature $T$, baryon number density
$n_B$ and proton-to-baryon ratio (or equivalently the electron
fraction) $Y_e$. Finally, at densities close to and above nuclear
matter density, the transition to a state of matter composed of
deconfined quarks may take place.

% Observations and statement of the problem
While accurate neutron star mass measurements of double pulsar systems
are plentiful, reliable radius measurements have only recently become
available. \citet{Steiner10} obtained quantitative constraints on the
EOS of nuclear matter from mass and radius measurements from quiescent
low-mass X-ray binaries and objects with photospheric radius expansion
bursts. Current neutron star radius observations suggest small radii
and lower pressures just above the nuclear saturation density. Small
neutron star radii could be interesting for the supernova dynamics,
because expects more gravitational binding energy. Unfortunately very
few available supernova EOS give (i) nuclear matter properties
consistent with those inferred from experiment, (ii) maximum neutron
star masses large enough to be consistent with the recent measurement
of a 1.97 solar mass neutron star in \citet{Demorest10}, and (iii)
masses and radii consistent with recent neutron star observations of
e.g.~\citet{Steiner10}. Furthermore, most available EOS tables are
based on interactions with larger values of the density derivative of
the nuclear symmetry energy, $L$, even though lower values are
suggested by both intermediate-energy heavy-ion
collisions~\citep{Tsang09}, chiral effective field
theory~\citep{Hebeler10b,Tews12} and neutron star
radii~\citep{Steiner12,Steiner12b}.

We should note that there are potential systematic uncertainties in
the neutron star mass and radius measurements which are not yet taken
into account. The relationship between the Eddington flux and the
point at which the photosphere returns to the neutron star surface is
not under control~\citep{Steiner10}, the evolution of the spectrum
during the tail of the burst is not well
understood~\citep{Suleimanov11}, and the value of the factor which
corrects for the fact that the X-ray spectrum is not a black body may
also modify inferred radii. Also, the X-ray spectrum may be modified
by accretion and violations of the assumed spherical symmetry. Recent
work~\citep{Steiner12b} finds that neutron star radii may be as large
as 13 km.

% Our solution
In this article, we construct new EOSs which are both consistent with
experimental nuclear data and recent neutron star mass observations.
We parameterize nucleonic matter with a new relativistic mean field
(RMF) model and describe nuclei and non-uniform nuclear matter with
the statistical model from \citet{Hempel:2009mc}. The final EOS is
also provided in tabular form, covering a wide range in density,
temperature, and electron fractions. We apply
these EOS tables in core-collapse SN simulations of intermediate- and
high-mass progenitors.

\section{Equations of state}

% Introduce the old models
The most commonly used EOSs are that of \citet{Lattimer:1991nc}
(hereafter LS), which is based on the compressible liquid-drop model
including surface effects, and H. \citet{Shen:1998gg} (hereafter
STOS), which is based on the TM1 RMF interaction~\citep{Sugahara94}
and uses the Thomas-Fermi approximation to describe non-uniform
nuclear matter. Both EOSs simplify the baryon composition using the
single-nucleus approximation (SNA) for heavy nuclei, and ignore all
light nuclei except for alpha particles. There are several studies of
the differences between these two EOSs in core-collapse simulations
\citep[see, e.g.,][]{Sumiyoshi:2006id,Sumiyoshi:2007pp,
  Fischer:2009,Connor:2011,Hempel:2012}. More recently, several new
EOS based on RMF interactions have become available. The new hadronic
EOS tables of G. Shen~\citep{shen2011a,shen2011b} are based on
NL3~\citep{Lalazissis97} and FSUgold~\citep{ToddRutel05} RMF
interactions with nuclei described in the Hartree appoximation. The
EOS model of \citet{Hempel:2009mc} and \citet{Hempel:2012} (hereafter
HS) is based on the statistical approach and is also used in the
present study. It goes beyond SNA by including the detailed nuclear
composition, based on experimentally measured nuclear masses as well
as different theoretical mass models. Tables are available for seven
different RMF parameterizations, including TM1, FSUGold, and
TMA~\citep{Toki95} and the new parameterizations SFHo and SFHx
developed in the present article.

% Mention the quark models
In addition to the hadronic EOS a first-order phase transition to
quark matter has been studied in \citet{Sagert:2008ka} and
\citet{Fischer:2011}, where it has been demonstrated that it can
trigger explosions even in spherically symmetric supernova models.
Note that the hybrid EOSs used result in extremely compact neutron
stars. All of these EOS studies in SN simulations explore also very
massive progenitor stars which in the end collapse to a black hole,
for which the post-bounce time until that moment can be used as an
observable to constrain characteristics of the phase transition.

% Review K, S, and L
There are several critical parameters for characterizing the equation
of state of hadronic matter, and some of the most relevant and yet
uncertain parameters are the nuclear incompressibility, $K$, the
symmetry energy at the saturation density, $J$, and the logarithmic
derivative of the symmetry energy $L$. The compressibility was an
important parameter for early core-collapse
simulations~\citep{Baron85}, and the compressibilities of the LS EOS
tables are available for 180, 200, and 375 MeV. However, the lowest
and highest of these values are far outside the currently acceptable
range of 240$\pm$10 MeV~\citep{Colo04}. We note that there is still
some model dependence in extracting this
value~\cite{Piekarewicz10,Khan12}. Early simulations also suggested
that the symmetry energy is important~\citep{Sumiyoshi95}. Recent
constraints on $J$ suggest $28<S<34$ from a combination of constraints
from experiments, theory, and observations of neutron star masses and
radii~\citep{Hebeler10b,Steiner12,Lattimer:2012,Tsang12,Steiner12b,Tews12}.
These works also imply constraints on $L$, but these constraints are
more model-dependent and are not always consistent with each other.
Only one of the original supernova EOS tables, the LS table with
$K=220$ MeV, obeys the current constraints on these EOS parameters.

\begin{table}
\caption{RMF Model parameters}
\begin{tabular}{llrlrl}
\hline
Quantity & Unit & SFHo & & SFHx & \\
\hline
$c_{\sigma}$ & fm & 3.1780 & & 3.4016 & \\
$c_{\omega}$ & fm & 2.2726 & & 2.5730 & \\
$c_{\rho}$ & fm & 2.4047 & & 2.4199 & \\
b & & 7.4653 & $\times 10^{-3}$ & 4.8157 & $\times 10^{-3}$ \\
c & & $-$4.0887 & $\times 10^{-3}$ & $-$4.3984 & $\times 10^{-3}$ \\
$\zeta$ & & $-$1.7013 & $\times 10^{-3}$ & 4.4218 & $\times 10^{-3}$ \\
$\xi$ & & 3.4525 & $\times 10^{-3}$ & 2.0535 & $\times 10^{-4}$ \\
$a_1$ & fm$^{-1}$ & $-$2.3016 & $\times 10^{-1}$ & 
$-$4.6241 & $\times 10^{-1}$ \\
$a_2$ & & 5.7972 & $\times 10^{-1}$ & 1.6604 & \\
$a_3$ & fm & 3.4446 & $\times 10^{-1}$ & 1.1792 & $\times 10^{-2}$ \\
$a_4$ & fm$^2$ & 3.4593 & & 2.1595 & $\times 10^{1}$ \\
$a_5$ & fm$^3$ & 1.3473 & & 1.5478 & \\
$a_6$ & fm$^4$ & 6.6061 & $\times 10^{-1}$ & 8.5506 & $\times 10^{-1}$ \\
$b_1$ & & 5.8729 & & 8.4606 & \\
$b_2$ & fm$^2$ & $-$1.6442 & & $-$2.3629 & \\
$b_3$ & fm$^4$ & 3.1464 & $\times 10^{2}$ & 4.0622 & $\times 10^{1}$ \\
$m_{\sigma}$ & fm$^{-1}$ & 2.3714 & & 2.3844 & \\
\hline
\end{tabular}
\label{tab:mod_param}
\end{table}

\begin{table*}[t]
\caption{Properties at saturation density and neutron star properties
  for the the different EOSs under investigation. The definition of
  all the quantities is given in the text.} \centering
\begin{tabular}{c c c c c c c c c c c c}
\hline
\hline
& $n_B^0$ & $E_{0}$ &$ K$ & $K'$ & $J$ & $L$ & 
$m_n^*/m_n$ & $m_p^*/m_p$ & R$_{1.4}$ & M$_{\text{T=0,Max}}$ 
& M$_{\text{s=4,Max}}$ \\
EOS & {[fm$^{-3}$]} & [MeV] & [MeV] & [MeV] &[MeV] & [MeV] & 
- & - & [km] & [M$_{\odot}$] & [M$_{\odot}$] \\
\hline
SFHo & 0.1583 & 16.19 & 245.4 & -467.8 & 31.57 & 47.10 & 
0.7609 & 0.7606 & 11.88 & 2.059 & 2.27 \\
SFHx & 0.1602 & 16.16 & 238.8 & -457.2 & 28.67 & 23.18 & 
0.7179 & 0.7174 & 11.97 & 2.130 & 2.36 \\
STOS(TM1) & 0.1452 & 16.26 & 281.2 & -285.3 & 36.89 & 110.79 & 
0.6344 & 0.6344 & 14.56 & 2.23 & 2.62 \\
HS(TM1) & 0.1455 & 16.31 & 281.6 & -286.5 & 36.95 & 110.99 & 
0.6343 & 0.6338 & 13.84 & 2.21 & 2.59 \\
HS(TMA) & 0.1472 & 16.03 & 318.2 & -572.2 & 30.66 & 90.14 & 
0.6352 & 0.6347 & 14.44 & 2.02 & 2.48 \\
HS(FSUgold) &  0.1482 & 16.27 & 229.5 & -523.9 & 32.56 & 60.43 & 
0.6107 & 0.6102 & 12.52 & 1.74 & 2.34 \\
LS(180) & 0.1550 & 16.00 & 180.0 & -450.7 & 28.61 & 73.82 & 
1 & 1 & 12.16 & 1.84 & 2.02 \\
LS(220) & 0.1550 & 16.00 & 220.0 & -411.2 & 28.61 & 73.82 & 
1 & 1 & 12.62 & 2.06 & 2.14 \\
\hline
\hline
\end{tabular}
\label{tab:exp_prop1}
\end{table*}

\begin{table}[t]
\caption{Properties of nuclei for SFHo and SFHx}
\centering
\begin{tabular}{c c c c c}
\hline
\hline
& E($^{208}$Pb) & R($^{208}$Pb) & E($^{90}$Zr) & R($^{90}$Zr) \\
& [MeV] & [fm] & [MeV] & [fm] \\
\hline
SFHo & -7.76 & 5.44 & -8.60 & 4.19 \\
SFHx & -7.87 & 5.41 & -8.55 & 4.19 \\
\hline
\hline
\end{tabular}
\label{tab:exp_prop2}
\end{table}

\begin{figure}
\includegraphics[width=3.4in]{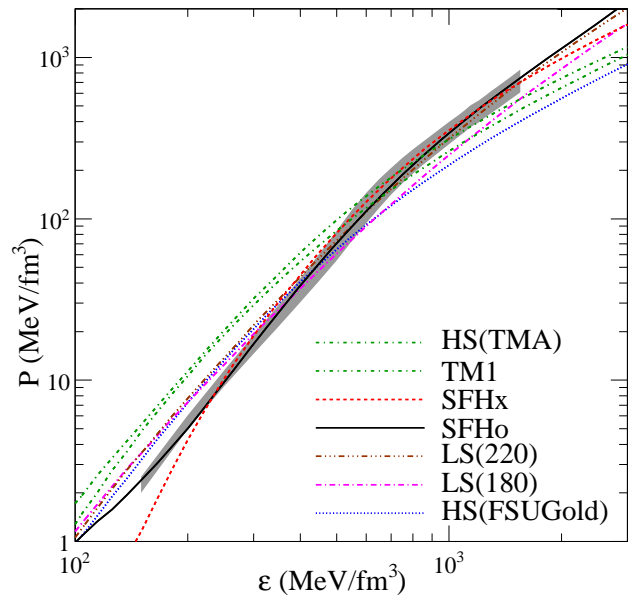}
\caption{Pressure of zero temperature beta-equilibrium matter as a
  function of the total energy density. The gray region gives the 1
  $\sigma$ confidence limits on the EOS from \citet{Steiner10}. The
  curve labeled TM1 applies for both the HS(TM1) and STOS EOSs. }
\label{fig:eos-t0}
\end{figure}

\begin{figure}
\includegraphics[width=3.4in]{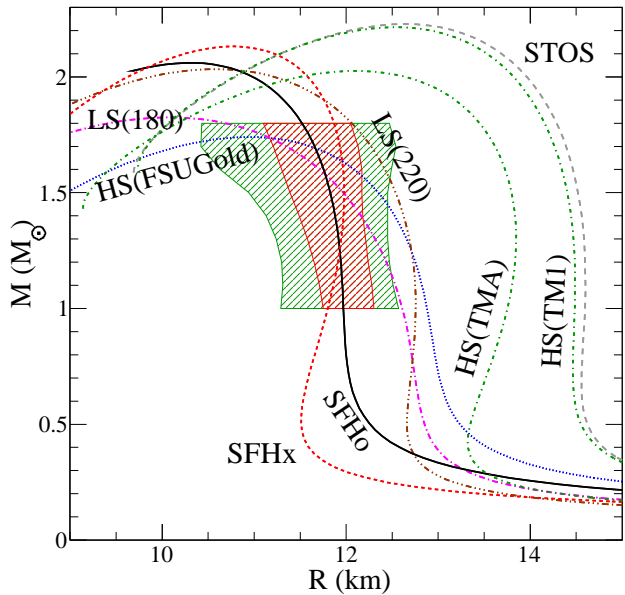}
\caption{Neutron star mass-radius curve for modern supernova equations
  of state. The red (green) region outlines the one (two) $\sigma$
  confidence limits from \citet{Steiner10}.}
\label{fig:mass-radius-t0}
\end{figure}

%Introduce our new EOSs: SFHo/SFHx
For the new EOS SFHo and SFHx, we choose to use a covariant Lagrangian
based on the Walecka model where nucleons interact via the exchange of
$\sigma$, $\omega$, and $\rho$ mesons in the mean-field approximation.
The non-linear Walecka model is well-known to have a limited range of
variation in the isospin sector and the nuclear symmetry energy is
controlled entirely through the coupling of the nucleons to the $\rho$
meson. In order to provide more flexibility, several additional terms
like $\rho^4$ and $\sigma^2 \rho^2$ have been considered. We use the
parameterization in \citet{Steiner05} which provides enough freedom to
modify the low- and high-density parts of the isospin sector
separately. There are 17 parameters including: the scalar-isoscalar
meson mass $m_{\sigma}$, the standard non-linear Walecka model
couplings $g_{\sigma}$, $g_{\omega}$, $g_{\rho}$, $b$, and $c$, the
two fourth-order vector meson couplings $\zeta$ and $\xi$, and the
parameters $a_{1-6}$ and $b_{1-3}$ which control the symmetry energy
as a function of density. These parameters are varied to ensure that
both EOSs have saturation properties which agree with that predicted
from nuclear masses and giant monopole resonances. In addition, both
predict binding energies and charge radii for $^{208}$Pb and $^{90}$Zr
that are within 2~\% of the experimental values. We ensure that the
pressure of neutron matter is always positive and always increases as
a function of the density. We ensure that the maximum mass is larger
than 1.93 solar masses (the lower $1-\sigma$ limit from
\citet{Demorest10}). We compare our results to LS(180) and FSUGold
even though these are inconsistent with this maximum mass because they
are still commonly used in the supernova community. The requirement
that the speed of sound is not superluminal is automatically enforced
in this fully covariant model. In our baseline model, SFHo, we also
fit the most probable mass-radius curve from \citet{Steiner10}, and in
our extreme model, SFHx, we attempt to minimize the radius of low-mass
neutron stars yet remaining consistent with the other constraints
given above. This forces the value of $L$ for SFHx to be on the lower
edge of the typical range of 20-120 MeV. We always take the mass of
the neutron to be 939.565346 MeV, the mass of the proton to be
938.272013 MeV, the mass of the $\omega$ meson to be 762.5 MeV, the
mass of the $\rho$ meson to be 770 MeV, and we use $\hbar c =
197.3269631$ MeV fm~\citep{Mohr08}. The full parameter list for both
models is given in Table~\ref{tab:mod_param}. The saturation
properties, nuclear binding energies and nuclear charge radii for our
models are given in Tables~\ref{tab:exp_prop1} and
\ref{tab:exp_prop2}. We remark that the values given in
Table~\ref{tab:exp_prop1} can be slightly different to previously
published ones, e.g.\ by \cite{hempel11b}, due to a different
treatment of the nucleon rest masses. In the present work we are using
the measured masses from above for all of the HS EOS, to obtain the
correct low-density limit. This treatment also leads to a slight
splitting of the neutron and proton effective masses, $m_n^*$ and
$m_p^*$ respectively. The value of $J$ for LS in
Table~\ref{tab:exp_prop1} differs from the published value by
\citet{Lattimer:1991nc} of 29.3~MeV. These authors computed $J$ as the
energy difference between neutron and nuclear matter whereas we are
calculating $J$ as the second derivative with respect to $Y_e$ at the
saturation point. In Table~\ref{tab:exp_prop1}, we also give the
saturation density $n_B^0$, the binding energy of nuclear matter
$E_0$, the skewness of nuclear matter $K^{\prime}$, the reduced
neutron and proton effective masses at saturation, the radii of 1.4
solar mass neutron stars at $T=0$, the maximum mass at $T=0$, and the
maximum mass for stars with constant entropy $s=4$ in beta-equilibrium
without neutrinos. The $T=0$ EOSs for all of the models constructed
with nucleon degrees of freedom are given in Figure \ref{fig:eos-t0},
along with the constraints from \citet{Steiner10}. Figure
\ref{fig:mass-radius-t0} gives the corresponding mass-versus radius
curves and also the constraints from \citet{Steiner10}.

% Talk about the low-density model
For densities below saturation density we apply the statistical model
from \citet{Hempel:2009mc} (HS) for the description of non-uniform
nuclear matter in nuclear statistical equilibrium (NSE), i.e.\ for the
formation of light and heavy nuclei within the gas of unbound
nucleons. For the unbound nucleons, we utilize the SFHo and SFHx
relativistic mean-field interactions. At low densities, the
description of nuclei is based on measured experimental binding
energies \citep{audi03}, which are combined with theoretical nuclear
structure calculations for exotic nuclei without measured data. Here
the finite range droplet model of \citet{Moller95} was chosen because
of its excellent reproduction of experimental binding energies, with a
rms deviation of only 0.669~MeV. Due to the use of nuclear structure
data, shell effects are automatically included. HS goes beyond the
single nucleus approximation and utilizes a distribution of different
nuclear species, and the results for light nuclei are in agreement
with more sophisticated quantum many-body models \citep{hempel11b}.
Also, the recent experimental study of \citet{qin2011} indicates that
the HS model is well suited for the description of matter at finite
temperature and densities around a few tenths of saturation density.
At even larger densities in the HS model the disappearance of nuclei
and smooth transition to uniform nuclear matter is assured by an
excluded volume description. Finally we calculate the EOS in tabular
form, covering densities from $10^{-12}$ to 10 fm$^{-3}$, temperatures
from 0.1 to 160 MeV, and electron fractions from 0 to 0.6, including
detailed information about the nuclear composition and the
thermodynamic properties. The tables are suitable for use in
astrophysical simulations and are available online.\footnote{See
  \texttt{http://phys-merger.physik.unibas.ch/\midtilde
    hempel/eos.html}\label{eospage}.}

% State which EOSs we will use in this paper
The resulting EOS will be compared below with the LS EOS, with
the different compressibilities 180~MeV (LS180), 220~MeV (LS220), and
375~MeV (LS375), and with STOS. Moreover, we will also include into
our comparison results obtained using the quark-hadron hybrid EOS from
\citet{Fischer:2011}. We select the model with bag constant
$B^{1/4}=155$~MeV and including corrections from the strong coupling
constant, $\alpha_S=0.3$ (hereafter QB155$\alpha_S$03), where the
phase transition to quark matter takes place at nuclear saturation
density for temperatures around 10~MeV and $Y_e\simeq0.3$ \citep[for
  details, see][]{Fischer:2011}. The hadronic part of this EOS table
is based on the STOS EOS, and these two EOS are identical at
sub-saturation densities where quarks are not present.

\section{Core-collape supernova simulations}

In this section, we will compare results from SN simulations obtained
using the SFHo EOS with the standard EOS LS180 and the two TM1 RMF
parameterizations STOS and HS. Furthermore, we will also compare SFHo
with the hybrid EOS QB155$\alpha_S$03, for which explosions were
obtained recently even in spherically symmetric
simulations~\citep{Fischer:2011}.
\begin{table}[htp]
\centering
\caption{Neutrino reactions considered including references.}
\begin{tabular}{cc}
\hline
\hline
Reaction\footnote{Note:
  $\nu=\{\nu_e,\bar{\nu}_e,\nu_{\mu/\tau},\bar{\nu}_{\mu/\tau}\}$ 
and $N=\{n,p\}$}
& References \\
\hline
$\nu_e + n \rightarrow p + e^-$ & \cite{Bruenn:1985en}\\
$\bar{\nu}_e + p \rightarrow n + e^+$ & \cite{Bruenn:1985en} \\
$\nu_e + (A,Z-1) \rightarrow (A,Z) + e^-$ & \cite{Langanke:2003ii}, \\
 & \citet{Hix:2003} \\
$\nu + N \rightarrow \nu' + N$ & \cite{Bruenn:1985en} \\
$\nu + (A,Z) \rightarrow \nu' + (A,Z)$ & \cite{Bruenn:1985en} \\
$\nu + e^\pm \rightarrow \nu' + e^\pm$ & \cite{Bruenn:1985en}, \\
 & \citet{Mezzacappa:1993gx}, \\
 & \citet{Mezzacappa:1993gm} \\
$\nu + \bar{\nu} \rightarrow e^- + e^+$ & \cite{Bruenn:1985en} \\
 & \citet{Mezzacappa:1999}\\
$\nu + \bar{\nu} + N + N \rightarrow N + N$ & \cite{Hannestad:1997gc} \\
$\nu_e + \bar\nu_e \rightarrow \nu_{\mu/\tau} + \bar\nu_{\mu/\tau}$
& \cite{Buras:2002wt} \\
\hline
\end{tabular}
\label{tab:nu-reactions}
\end{table}

\subsection{Supernova model}

% Agile-Boltztran SN model
Our core-collapse SN model, AGILE-BOLTZTRAN, is based on general
relativistic radiation hydrodynamics in spherical symmetry. It employs
three-flavor Boltzmann neutrino transport \citep[see][and references
  therein]{Liebendoerfer:2004}. We use the standard weak processes
following \citet{Bruenn:1985en}, see Table~\ref{tab:nu-reactions} for
details. In addition, we include the improved rates for
electron-captures on heavy nuclei from \citet{Langanke:2003ii} and
\citet{Hix:2003}, weak magnetism and nucleon recoil based on
\citet{Horowitz:2001xf}, and the annihilation of trapped electron
neutrino pairs has been implemented in \citet{Fischer:2009} following
\citet{Buras:2002wt}.

For NSE conditions ($T>0.45$~MeV), we implement the baryon EOS tables
specified above. For non-NSE, we assume the ideal gas of $^{28}$Si for
the baryon EOS. On top of the baryons, also for NSE, contributions
from electrons, positrons and photons are added to the EOS using
\citet{Timmes:1999}. Recently, this Si-gas approximation has been
replaced by a nuclear reaction network, based on the nuclear
composition given by the progenitor model. It allows, e.g., for a
smooth NSE-to-non-NSE transition as well as to simulate a large domain
of the progenitor star \citep[for details, see][]{Fischer:2009af}.

The simulations we will discuss further below are launched from
iron-core progenitors. We use the 15~M$_\odot$ model from
\citet{Woosley:2002zz} for regular core-collapse supernovae and the
40~M$_\odot$ from \citet{Woosley:1995ip} for the black-hole formation
scenario. None of the spherically symmetric simulations using the
purely hadronic EOS results in an explosion for the considered
simulation times.

\subsection{Simulation results}

A detailed supernova EOS comparison study for LS180 and STOS as well
as several HS EOS tables with different RMF parameterizations has
been published very recently in \citet{Hempel:2012}. Here, we extend
their analysis and include in addition SFHo/SFHx.

\begin{figure}
\includegraphics[width=\columnwidth]{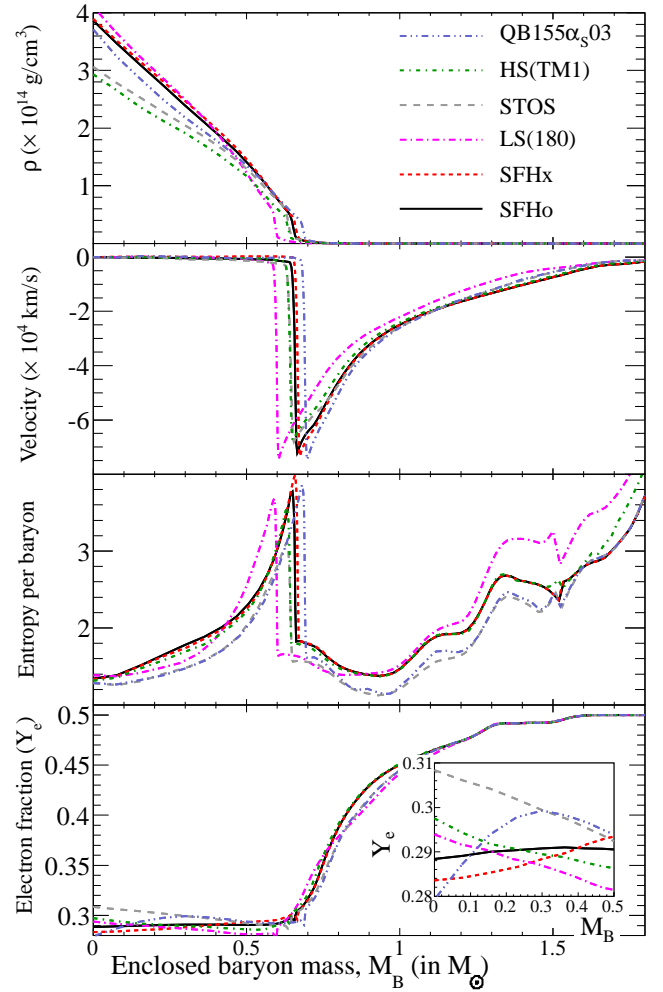}
\caption{Bounce profiles of selected quantities for the 15~M$_\odot$
models, comparing the different EOS under investigation. }
\label{fig:s15.0-bounce}
\end{figure}

% core collapse
During the core-collapse phase, the composition is dominated by heavy
nuclei. At low temperatures in non-NSE the EOS based on the ideal gas
of Si-nuclei has been applied in all simulations except for HS(TM1)
where we used our simplified nuclear reaction network. Hence, all EOS
lead to very similar structures in the simulations at low
temperatures, see e.g. the equal entropy per baryon profiles in the
outer layers ($M_{\text{B}}>1.5~$M$_\odot$) as illustrated in
Fig.~\ref{fig:s15.0-bounce} (except HS(TM1)). Note also that LS(180)
has a different non-NSE treatment based on the ideal Si-gas. Moreover,
structure differences arise at higher temperatures in NSE, which will
be discussed i+n the following paragraph. In general, low-density
differences (i.e. on top of the bounce shock
$0.7<M_{\text{B}}<1.5~$M$_\odot$ in Fig.~\ref{fig:s15.0-bounce})
between STOS and HS(TM1) are related to a different description of
heavy nuclei. The statistical approach of HS is largely based on
experimentally known masses including shell effects. It resembles the
ideal gas of $^{56}$Ni/$^{56}$Fe at the transition to non-NSE by
construction. In contrast, the Thomas-Fermi approximation of STOS
gives heavy nuclei which are too strongly bound and does not perform
well at this transition. \citet{Hempel:2012} explained how these
differences affect the $Y_e$ evolution. Furthermore, entropy
differences between STOS and HS(TM1) originate from the missing
kinetic entropy contribution of heavy nuclei in STOS. Since we use the
same low-density description for nuclei as in HS, simulations using
SFHo and SFHx result in the same conditions during the core-collapse
phase. The simulation results for QB155$\alpha_S$03 are similar to
STOS at low densities where quarks are not present. Even though they
are based on the same low-density EOS, small differences
have emerged because the core has already entered the quark-hadron
mixed phase.

%

% core bounce
Only a few tens of milliseconds before core bounce, nuclear saturation
density is reached at the center. At these conditions high-density EOS
differences become large which in turn lead to different dynamical
evolutions. We define the moment of core bounce when the central
density reaches its maximum before shock breakout. The EOS with
highest central density, LS(180), results in the smallest enclosed
baryon mass (see Fig.~\ref{fig:s15.0-bounce}). The two TM1 EOSs, STOS
and HS(TM1), result in the lowest central densities. Differences
between STOS and HS(TM1) are due to the different description of heavy
nuclei, as already discussed above. Furthermore, the presence of light
nuclei, which are taken into account in HS, also has an impact on
thermodynamic properties. Comparing STOS and QB155$\alpha_S$03, the
most pronounced differences arise above nuclear saturation densities
which are due to the presence of deconfined quarks (in particular
s-quarks). They soften the high-density EOS and result in higher
central densities and significantly lower central $Y_e$ at core bounce
\citep[for details, see][]{Fischer:2011}.

One of our new models, SFHx, gives the lowest electron fractions for
$M< 0.25$~M$_\odot$, due to its low slope of the symmetry energy $L$.
Compared with LS(180), SFHo and SFHx reach similarly high central
densities and low central $Y_e$ at core bounce (see
Fig.~\ref{fig:s15.0-bounce} and further discussion below). Note that
LS(180) has lowest incompressibility $K$ and symmetry energy $J$ among
all EOS under consideration. However, the baryon mass enclosed inside
the bounce shock for the two new EOS differs by about 0.1~M$_\odot$
and is actually more similar to the TM1 results, which has an
extremely high $K$ and $J$. These results provide no clear
correlations between individual nuclear matter properties of the EOS
(see Table~\ref{tab:exp_prop2}) and the different conditions, e.g.,
central density and $Y_e$ obtained at core bounce for these EOS (see
Fig.~\ref{fig:s15.0-bounce}). It seems difficult to disentangle
individual nuclear EOS properties from the different conditions
obtained in simulations of stellar core collapse.
%

\begin{figure}[t!]
\includegraphics[width=\columnwidth]{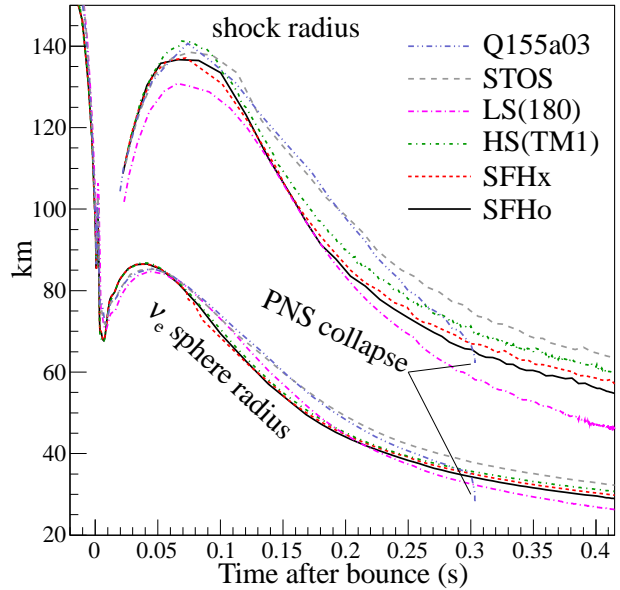}
\caption{Evolution of the shock radii (thick lines)
and $\nu_e$-spheres (thin lines) for the 15~M$_\odot$ models,
comparing the different EOS under investigation.}
\label{fig:shock}
\end{figure}

% post-bounce phase
The central object formed at core bounce, i.e. the protoneutron star
(PNS), is hot and lepton rich in which sense it differs from the final
remnant neutron star. After bounce, the shock starts to propagate
outwards with initially positive velocities. Simultaneously, mass
accretion from the outer part of the stellar core continuously grows
the mass of the PNS. Moreover, the expanding bounce shock loses energy
due to heavy-nuclei dissociation and neutrino emission. The neutrino
emission is related to a large number of electron captures on protons
during the shock passage across the neutrinospheres. It releases a
burst of $\nu_e$ of several $10^{53}$~erg~s$^{-1}$ for a short
timescale between 5--20~ms after core bounce. Both sources of energy
loss turn the expanding and dynamic (i.e. accompanied with matter
outflow) bounce shock into an accretion front, the standing accretion
shock (SAS), already between 5--10~ms after core bounce. The later PNS
evolution is determined from mass accretion and the subsequent PNS
compression which leads to continuously rising central density and
temperature. For a given mass accretion rate, determined from the
progenitor, and otherwise identical simulation setup, the timescale
for the PNS compression is directly related to the EOS. On timescales
on the order of several 100~ms, EOS differences lead to different
neutrinospheres and shock radii, displayed in Fig.~\ref{fig:shock}.
The neutrinospheres can be used to characterize the size of the
central protoneutron star as they are located in its outer envelope.

At 100~ms post bounce, LS(180) has the smallest shock radius at 130~km
because the shock had lower energies at bounce. However, its
contraction proceeds initially, up to about 200~ms post bounce,
somewhat slower than those for the EOS HS(TM1) and SFHo/SFHx. This
behavior is related to the rapid PNS contraction obtained for the
latter. By comparing with STOS, one can identify the effect of the
different treatment of heavy nuclei and in addition the consistent
inclusion of light nuclei for HS and also for SFHo/SFHx \citep[for
  details, see][]{Hempel:2012}. These differences of the low-density
EOS result in an initially accelerated PNS contraction of HS(TM1).
STOS leads to the generally slowest PNS contraction and hence slowest
shock contraction as well. Differences between STOS and
QB155$\alpha_S$03 are related to the presence of quarks, in particular
strange quarks, at and above saturation densities. These become
abundant during the PNS contraction as the PNS domain occupied by
quark matter grows continuously during the accretion phase \citep[for
  details about the scenario, see][]{Fischer:2011}. The high-density
EOS is dominated by the quark matter description after about 200~ms
post bounce. This results in an accelerated contraction of the PNS for
QB155$\alpha_S$03 in comparison to STOS (see Fig.~\ref{fig:shock}).
Note that at about 300~ms post bounce, the PNS becomes gravitationally
unstable and collapses. A massive quark core forms at the PNS center,
which causes the collapse to halt and an additional shock wave forms
at saturation density. The subsequent shock expansion and even
acceleration, accompanied by mass outflow, determines the onset of
explosion for this model. Since none of the purely hadronic EOSs lead
to an explosion, we only show and discuss results up to the moment of
PNS collapse for QB155$\alpha_S$03. 

\begin{figure}[t!]
\includegraphics[width=\columnwidth]{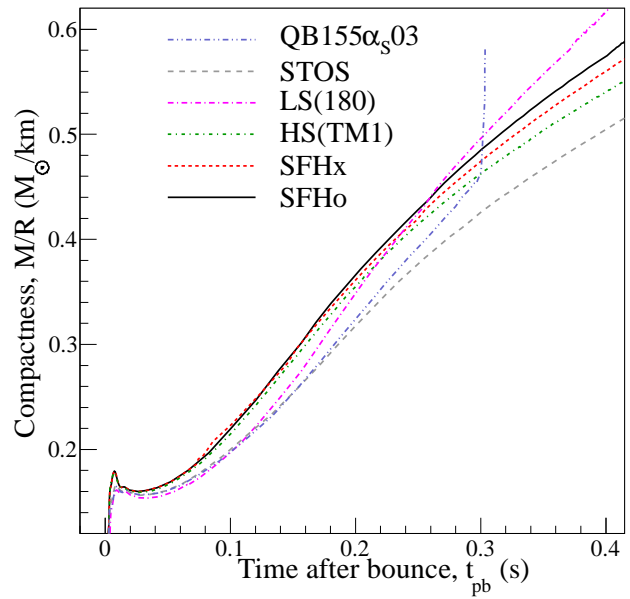}
\caption{The compactness of the protoneutron star after bounce,
defined by the gravitational enclosed mass and radius of the
electron neutrinosphere.}
\label{fig:compactness}
\end{figure}

It is difficult to predict astrophysical simulation results from only
nuclear matter properties at saturation density and zero temperature.
For example both HS(TM1) and STOS are based on the same nucleon
interactions TM1 (see Table~\ref{tab:exp_prop2}). However, the
additional inclusion of light nuclei and the different description of
heavy nuclei, relevant for the low- and intermediate-density EOS have
a significant impact on the conditions at bounce and then also on the
later PNS structure and contraction behavior. The early PNS evolution
for our new parameterizations SFHo and SFHx, up to about 50~ms post
bounce, resembles those of HS(TM1) (see the neutrinospheres in
Fig.~\ref{fig:shock}). Later, the PNS contraction proceeds even faster
than for HS(TM1), reflecting the softer high-density EOS of SFHo/SFHx.
On timescales of several 100~ms, the PNS and shock contractions slow
down for all the RMF EOSs as a consequence of the continuously growing
enclosed mass. This effect is significantly weaker for the
non-relativistic LS(180) EOS.

Fig.~\ref{fig:compactness} shows the evolution of the compactness, the
ratio of the enclosed gravitational mass to the radius. We evaluate
these quantities at the electron neutrinosphere, representative of the
protoneutron star surface. The EOSs which result in the most compact
protoneutron stars at core bounce and few tens of milliseconds after
are those of SFHo and SFHx, closely followed by HS(TM1). At core
bounce and up to 50~ms post bounce, LS180 results in the least compact
protoneutron star. However, during the later post-bounce evolution for
LS180 the compactness of the central protoneutron star grows fastest
among all of the EOSs under exploration. For HS(TM1) the protoneutron
star contraction proceeds on the same timescale as STOS, only on a
slightly more compact magnitude. This can be attributed to the more
compact, configuration at core bounce for HS(TM1). STOS
and QB155$\alpha_S$03 proceed along identical lines until about 100~ms
post bounce, when quark matter begins to dominate the protoneutron
star evolution. Note, at that moment a quark-hadron phase transition
to strange quark matter has not been fully achieved, i.e. most mass
inside the protoneutron star is only in the quark-hadron mixed phase.
The EOSs with most compact protoneutron stars at bounce and during the
early post-bounce evolution are SFHo and SFHx which remain the most
compact ones until about 300~ms post bounce, when only LS180 becomes
more compact. Up to this moment, SFHo and SFHx have the same
compactness, and differences observed during the later evolution
remain small.

% Protoneutron star collapse and black hole formation
In addition to the simulations using the 15~M$_\odot$ progenitor, we
also include the more massive 40~M$_\odot$ progenitor from
\citet{Woosley:2002zz} into our EOS comparison in supernova
simulations. This progenitor has been discussed before within the
scenario of black-hole formation, also explored for the LS and STOS
supernova EOSs \citep[see,
  e.g.,][]{Sumiyoshi:2006id,Sumiyoshi:2007pp,Nakazato08,Fischer:2009,Connor:2011,Hempel:2012}.
The moment of black hole formation is defined, within our general
relativistic framework \citep[for details, see][]{Liebendoerfer:2004},
as the central lapse function approaches zero and stable numerical
solutions for the evolution equations cannot be obtained anymore. The
lapse function determines, e.g., time dilatations and gravitational
Doppler shift effects. Note that core-collapse events which result in
a black hole instead of an explosion are as bright as normal supernova
explosions with respect to neutrino signal. The possible future
detection of such a neutrino signal from a Galactic event may be used
to further constrain the high-density EOS, see \citet{nakazato2010}.

\begin{figure}[htp]
\includegraphics[width=\columnwidth]{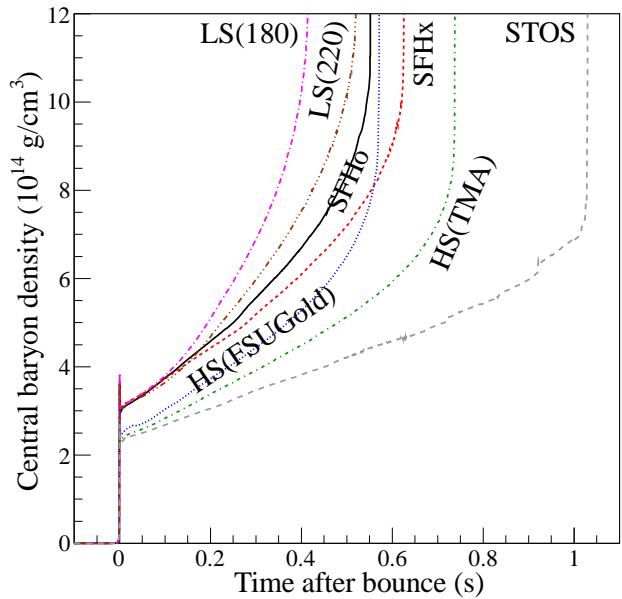}
\caption{Central density evolution for the 40~M$_\odot$ model,
comparing the hadronic EOS under investigation.}
\label{fig:density}
\end{figure}

The moment of black-hole formation is approached during the
post-bounce evolution, during which the mass enclosed inside the PNS
grows continuously in the absence of an explosion. The timescale for
the central density (and also temperature) rise is given by the PNS
contraction behavior, which in turn depends on the nuclear EOS. We
list selected properties at the onset of PNS collapse in
Table~\ref{tab:bh} for all hadronic EOS under investigation.

The PNS compression can be illustrated best via the central density
evolution, shown in Fig.~\ref{fig:density}, comparing the different
hadronic EOSs under investigation. The values of the central density
cluster around only two different values, 3.2$\times 10^{14}$ and
2.5$\times 10^{14}$ g/cm$^{3}$, immediately after the bounce. It is
likely that this is related to the properties of the EOS at
low-density, and thus also to the radii of very low-mass neutron
stars. A similar (but not exact) clustering is observed in the
lower-right hand corner of Fig.~\ref{fig:mass-radius-t0} where STOS,
HS(TM1), and HS(FSUGold) have larger radii for 0.4 solar mass neutron
stars. This grouping lasts only for 300 ms after which time the EOS at
higher densities becomes important.

HS(TMA) has a higher incompressibility but lower symmetry energy than
STOS (which uses TM1) and leads to a significantly shorter (about
300~ms) accretion time until black-hole formation (see
Fig.~\ref{fig:density}). This would indicate that the symmetry energy
plays the dominant role. However, SFHo and SFHx do not obey this
trend. Even though SFHo has a higher incompressibility and higher
symmetry energy than SFHx, it leads to a shorter (about 50~ms)
accretion time until black-hole formation. Instead of using nuclear
matter properties, one could expect that our results could be
explained by the maximum mass of cold neutron stars, which is an
integrated quantity of the EOS. However, by comparing the times until
black hole formation from Table~\ref{tab:bh}) with the numbers in
Table~\ref{tab:exp_prop1}, it turns out that there is no monotonic
correlation. For example the collapse of FSUgold occurs about 160~ms
later than for LS(180), even though the maximum mass of LS(180) 
is 0.1~M$_\odot$ larger.

\begin{figure}[htp]
\includegraphics[width=\columnwidth]{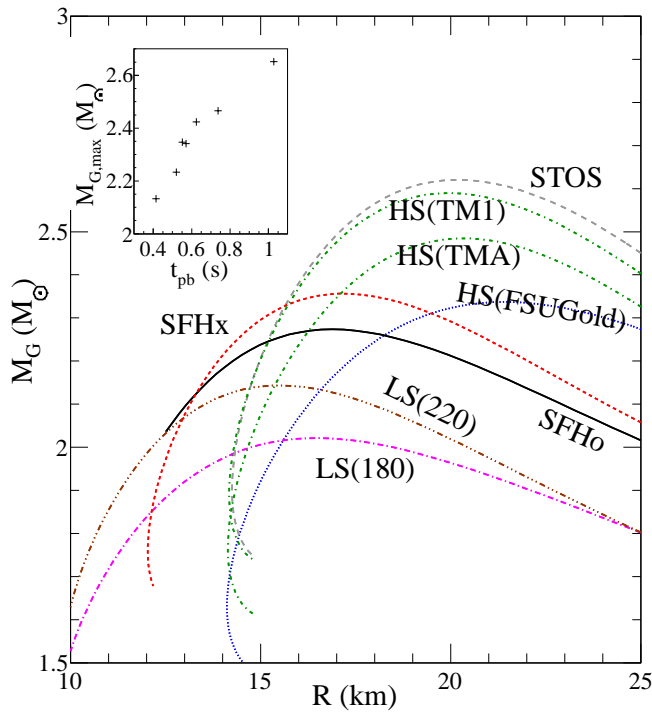}
\caption{Mass-radius relations at constant entropy per baryon of
  4~k$_B$ comparing the hadronic EOSs under investigation.}
\label{fig:mass-radius-s4}
\end{figure}

Note that the two RMF EOS parameterizations, TM1 and TMA, fail to
fulfill both zero temperature maximum mass and radius constraints,
while LS(220) is on the edge of acceptance, see
Fig.~\ref{fig:mass-radius-t0}. The RMF EOS based on TM1 and TMA result
in sufficiently high maximum masses but give at the same time radii
which are too large. With the black-hole formation scenario explored
here, these large radii are due to slower
PNS contraction which results in an extended post-bounce mass
accretion period until black-hole formation (see
Fig.~\ref{fig:density}). On the other hand, FSUgold results in
reasonably small neutron star radii but fails to fulfill the current
maximum zero temperature mass constrain of $1.97\pm0.04$~M$_\odot$. It
leads to a significantly shorter time until black-hole formation than
for TMA and TM1, comparable to SFHo which has a larger
incompressibility but slightly lower symmetry energy than FSUgold. On
the other hand, SFHo and SFHx fulfill both mass and radius constraints
by construction. It is interesting that the models which fulfill the
mass-radius and maximum mass constraints, i.e.\ LS(220), SFHo, and
SFHx, give rather similar results for $t_{BH}$, which seems to be
constrained to $500 - 650$~ms. It appears that neutron star radius
measurements constrain the time to black hole formation for 40
$M_{\odot}$ progenitors.

\begin{table}
\caption{Selected quantities at the onset of PNS collapse and
  black-hole formation.} \centering
\begin{tabular}{l c c c c c}
\hline
\hline
EOS &
$t_{pb}$ \footnote{time post bounce} &
$\rho$ \footnote{central baryon density} &
$T$ \footnote{central temperature} &
$M_B$ \footnote{enclosed baryon mass} &
$M_G$ \footnote{enclosed gravitational mass}
\\
& $[s]$ & $[10^{15}$ g/cm$^3 ]$ & $[MeV]$ & $[$M$_\odot]$ & $[$M$_\odot]$
\\
\hline
LS(180)     & 0.415 & 1.292 & 29.978 & 2.227 & 2.133 \\
LS(220)     & 0.521 & 1.324 & 31.446 & 2.350 & 2.233 \\
SFHo         & 0.551 & 1.067 & 46.691 & 2.477 & 2.347 \\
HS(FSUgold) & 0.571 & 1.058 & 48.104 & 2.465 & 2.341 \\
SFHx        & 0.625 & 0.803 & 40.830 & 2.552 & 2.424 \\
HS(TMA)     & 0.737 & 0.943 & 46.708 & 2.626 & 2.466 \\
STOS         & 1.028 & 0.769 & 49.705 & 2.864 & 2.652 \\
\hline
\hline
\end{tabular}
\flushleft
{\bf Notes:}
\label{tab:bh}
\end{table}
\citet{Hempel:2012} demonstrated that maximum neutron star mass
determined from the EOS in beta-equilibrium at $s=4$~k$_B$/baryon was
strongly correlated with the time until black-hole formation. This
time can be measured with currently available neutrino detectors given
a galactic core-collapse supernova. The associated mass versus
radius curves and their relationship with the time to black hole
formation is displayed in Fig.~\ref{fig:mass-radius-s4} for the
hadronic EOSs under investigation. The corresponding values are also
listed in the last column of Table~\ref{tab:exp_prop1}. Note that the
maximum masses increase for all EOS compared to the $T=0$ case.
(compare also Figs.~\ref{fig:mass-radius-s4} and
\ref{fig:mass-radius-t0}). Pressure is more sensitive than the energy
density to the temperature effects because of the large nucleon mass
and thus larger temperatures in increase the maximum mass. If we
compare the maximum masses of the $s=4$~k$_B$/baryon configurations from
Table~\ref{tab:exp_prop1} with the time until black-hole formation
from Table~\ref{tab:bh}, we find that these quantities have the same
ordering, i.e. show an (almost) strictly monotonic correlation. 

\section{Summary and conclusions}

We developed two new relativistic mean-field (RMF) interactions, based
on which we constructed the two new supernova EOSs, SFHo and SFHx, in
such a way to fulfill current constraints from neutron star mass and
radius measurements. Moreover, the new EOSs are consistent with
nuclear experimental constraints on matter near and below the
saturation density. These new EOSs provide more variation in the set
of EOS tables which can be used by the core-collapse supernova
community instead of EOSs based on TMA or TM1 which are now ruled out
by observations.

The new EOS were implemented in core-collapse supernova simulations of
massive iron-core progenitors. We compared the results with the
commonly used non-relativistic EOS of \citet{Lattimer:1991nc} and the
RMF EOS of \citet{Shen:1998gg}. Moreover, we include into our EOS
comparison also the recently introduced quark-hadron hybrid EOS from
\citet{Fischer:2011} and the RMF EOS from \citet{Hempel:2009mc} based
on the parameterizations TM1, TMA and FSUgold. We compared the
different EOS during the iron-core collapse phase, which is dominated
by heavy nuclei, and confirmed already reported differences between
these EOS \citep[see, e.g.,][]{Hempel:2012}. We extend the analysis
and include SFHo/SFHx. Differences become large only slightly before
and after core bounce, when the central density exceeds normal nuclear
matter density. The post-bounce mass accretion phase is ideal to study
the protoneutron star contraction behavior, which reflects the EOS
underlying nuclear matter properties for a given progenitor choice. We
found that the two new EOS which give small radii for cold neutron
stars also lead to the most compact protoneutron stars in the first
300~ms after bounce. However, it is not easy to disentangle the
relationship between individual nuclear matter properties, given at
zero temperature near the saturation density, and the outcome of our
supernova simulations. 

We also find the EOS classifications 'soft' or 'stiff' misleading.
Implicitly what is meant by soft or stiff is that the EOS has a lower
or higher pressure. However, core-collapse supernova explore a large
range of densities and temperatures, and an EOS which has a higher
pressure at one density and temperature may have a lower pressure at
another density and temperature. This is particularly evident with the
black-hole formation time as described here and in
\citet{Hempel:2012}, as EOSs like SFHx which have a low pressure at
zero temperatures near the saturation density, have a larger time
until black hole formation than the other EOSs because their pressure
at $s=4$ is larger. The addition of quark degrees of freedom only
further complicates this issue.

Moreover, we explored the possible correlation between the EOS and
outcome of the black-hole formation scenario in the absence of a
supernova explosion. We confirm the analysis of \citet{Hempel:2012},
where only a correlation between finite entropy per baryon
configurations of static protoneutron stars and the time until
black-hole formation was found. Note that the neutrino signals of
these events are as bright as ordinary supernova explosions. But at
the moment of black-hole formation the signal suddenly stops which
makes this time measurable with neutrino telescopes. The possible
future neutrino observation of such a Galactic event will constrain
the high-density and finite-entropy EOS significantly, complementary
to future neutron star mass and radius observations.

One conclusion from this study is that it seems to be difficult to
understand the effect of the EOS in core-collapse supernovae by
analyzing non-exploding models. To tackle the impact of the EOS on the
explosion mechanism, multi-dimensional simulations are necessary, see
e.g.\ \citet{marek09}. Furthermore, it remains to be explored how
different nuclear EOS can influence long-term cooling to protoneutron
stars after the supernova explosion has been launched. Therefore, the
implementation of weak processes consistent with the nuclear EOS is
required \citep[see, e.g.,][]{Reddy:1998,Reddy:1999}. This may impact
the neutrino cooling timescale and also the extent of protoneutron
star convection as studies recently described by \citet{Roberts:2012}.
The nuclear EOS may also be important for the proton-to-baryon ratio
of the material ejected form the protoneutron star surface known as
neutrino-driven wind relevant for the nucleosynthesis of heavy
elements. Moreover, a possible EOS impact within the cooling of
protoneutron stars on the emitted neutrino signal may also be of
relevance for neutrino oscillation studies, in particular those which
explore collective phenomena in the presence of large neutrino but
small matter densities.

\section*{Acknowledgements}

The authors would like to thank S. Reddy, L. Roberts and I. Sagert for
several discussions on this work. The supernova simulations were
performed at the computer center of the Helmholtzzentrum f\"ur
Schwerionenforschung GmbH (GSI) in Darmstadt, Germany. A.W.S. was
supported by DOE grant DE-FG02-00ER41132. M.H. and T.F. are grateful
for support from CompStar, a research networking program of the ESF.
M.H.\ acknowledges support from the High Performance and High
Productivity Computing (HP2C) project, the Swiss National Science
Foundation (SNF) under project number no. 200020-132816/1, and
ENSAR/THEXO. T.F. is supported by the Swiss National Science
Foundation under project~no.~PBBSP2-133378 and HIC for FAIR.

\bibliographystyle{apj}
\bibliography{paper}

\end{document}